\documentclass{nature}
\usepackage{color}
\usepackage[utf8]{inputenc}
\usepackage{amsmath, amssymb}
\usepackage{txfonts}
\usepackage{xspace}
\usepackage{graphicx}

\newcommand{\quot}[1]{``#1''}
\newcommand{\VEC}[1]{\mathbf{#1}}
\newcommand{\phyp}{p_R}
\newcommand{\ppair}{\rho_r}
\newcommand{\SI}[1]{Supplementary Discussion~#1\xspace}
\newcommand{\eq}[1]{Eq.~(\ref{e:#1})}
\newcommand{\elabel}[1]{\label{e:#1}}
\newcommand{\flabel}[1]{\label{f:#1}}
\newcommand{\fig}[1]{Fig.~\ref{f:#1}}
\let\includegraphicsn\includegraphics

\newcommand{\xvec}{\VEC{x}}
\newcommand{\pvec}{\VEC{p}}

\title{Efimov-driven phase transitions of the unitary Bose gas}
\author{Swann Piatecki$^1$ \& Werner Krauth$^1$} 
\begin{document}

\maketitle
\begin{affiliations}
  \item Laboratoire de Physique Statistique, École Normale Supérieure,
UPMC, Université Paris Diderot, CNRS, 24 rue Lhomond, 75005 Paris, France
\end{affiliations}
\maketitle
\begin{abstract}
In quantum physics, Efimov trimers\cite{Efimov1970, Efimov1971, 
Efimov1979} are bound states of three 
particles that fall apart like Borromean rings when one
of them is removed. Initially predicted in nuclear physics, 
these striking bosonic states are hard to observe, but the \quot{unitary}
interactions at which they form is commonly realized in current
cold atoms experiments\cite{Gross2009,
Pollack2009, Zaccanti2009, Rem2013}. There,
they set the stage for a new class of universal physics:
Two-body interactions are all but invisible, but three-body effects allow the 
emergence of a largely uncharted new world of many-particle bound states. 
Three-particle systems
were characterized theoretically\cite{Braaten2006}, and
the ground-state properties of small unitary clusters
computed numerically\cite{Stecher2010}, but the macroscopic many-body behaviour has remained
unknown. Here we show, using
a Path-Integral Monte Carlo algorithm\cite{Pollock1984,
Krauth1996, Navon2011} backed up by theoretical
arguments, that the unitary Bose gas presents a first-order phase transition 
from a normal gas to a superfluid Efimov liquid. 
The normal gas is very well described by the
available virial coefficients\cite{Castin2013}. 
At unitarity, the phase diagram of the bosonic system is universal
in rescaled pressure and temperature.
A triple point separates
the normal gas, the superfluid Efimov liquid, and a third phase, the
conventional superfluid gas. These two superfluid phases are separated
by a critical line that ends in a critical point at high temperature.
This rich phase diagram should allow for a number of experimental protocols that
would probe these universal transitions between the normal gas,
the superfluid gas, and the superfluid Efimov liquid.

\end{abstract}

We describe the system of $N$ interacting bosons by a hamiltonian
\begin{equation}
  H=\sum_{i}\frac{\pvec_i^2}{2m} + \sum_{i< j}V_2^a(r_{ij}) + 
  \sum_{i < j < k}V_3(R_{ijk}) +  \frac{m \omega^2}{2}\sum_i\xvec_i^2, 
  \elabel{hamiltonian}
\end{equation}
where $\pvec_i$, $\xvec_i$ and $m$ are the momentum, the position and the mass
of particle $i$. The pair interaction $V_2^a$ is of range zero, and
its unitary point corresponds to an infinite scattering length~$a$.
The three-body interaction $V_3$ implements a hard-core hyperradial 
cutoff condition, $R_{ijk} > R_0$, 
where the hyperradius $R_{ijk}$ of particles $i$, $j$, and $k$ 
corresponds to their root-mean-square pair distance ($3R_{ijk}^2 = 
r_{ij}^2 + r_{ik}^2 + r_{jk}^2$). This three-body hard core prevents the so-called Thomas 
collapse\cite{Thomas1935} into a many-body state with
vanishing extension and infinite negative energy by setting a fundamental 
trimer energy $-E_t\propto R_0^{-2}$. The final term in \eq{hamiltonian} 
models an isotropic harmonic trapping potential of length $a_\omega=\sqrt{\hbar/(m\omega)}$ as
it is realised in ultracold bosons experiments (see \SI{1}
for a full description of the hamiltonian).  
The properties of the system described by \eq{hamiltonian} are universal
when $R_0$ is much smaller than all other length scales. Unlike the hamiltonian
of \eq{hamiltonian}, experimental
systems of cold atoms tuned to unitarity at a Feshbach 
resonance\cite{Chin2010} are metastable because of the presence of deeply bound states. When
three particles get too close, they leave the trap
and contribute to the notorious three-body losses. In experiments, the
three-body parameter given by $R_0$ depends on pair interaction parameters specific to each
atomic species\cite{Ferlaino2011}.

In thermal equilibrium, within the path-integral representation of quantum
systems that we use for our 
computations, position variables $\xvec_i(\tau)$ carry an imaginary-time index
$\tau \in [0,\beta= 1/k_B T]$, where $k_B$ is the Boltzmann constant. The
fluctuations of $\xvec_i(\tau)$ along $\tau$ account for the quantum
uncertainty. The bosonic nature of many-particle systems manifests itself through
the periodic boundary conditions in $\tau$ and, in particular, through the
permutation structure of particles. The length of permutation cycles 
correlates with the degree of quantum coherence\cite{Feynman1982, Krauth2006}.
Interactions set the statistical weights of configurations\cite{Pollock1984}.
Ensemble averaging, performed by a dedicated Quantum Monte Carlo algorithm,
yields the complete thermodynamics of the system (see \SI{2} for computational
details). The $N$-body simulation code is
massively run on a cluster of independent processors. It succeeds
in equilibrating samples with up to a few hundred bosons. 

Snapshots for three particles in a shallow trap at $\omega \sim 0$ 
illustrate the quantum fluctuations in $\xvec_i(\tau)$, and characterize
the Efimov trimer.
Indeed, for two-body interactions without a bound state,
virtually free particles fluctuate on the scale of the de~Broglie wavelength 
$\lambda_\text{th}=\sqrt{2\pi\hbar^2\beta/m}$ that diverges at low temperature (see \fig{snapshot}{\bf a}). 
In contrast, for positive $a$, a bound state with energy 
$-E_\text{dimer}=-\hbar^2/(ma^2)$ forms in the
two-body interaction potential. Two particles  bind into a
dimer, and the third particle is free (see \fig{snapshot}{\bf c}). 
Efimov physics takes place
at unitarity, where the bound state of the pair
potential is at resonance $E_\text{dimer} = 0$, and the scattering length $a$ is
infinite (see \fig{snapshot}{\bf b})). At this
point, the two-body interaction is scale-free. 
While two isolated particles do not bind, in the three-particle
system, pairs of particles approach each other, and then dissociate, so that, between
$\tau =0$ and $\tau = \beta$, the identity of close-by partners changes several
times. This coherent particle-pair scattering process, the hallmark of the
Efimov effect\cite{Efimov1979}, is highlighted in 
\fig{snapshot}{\bf b.}  At small temperatures, the fluctuations of this
bound state remain on a scale proportional to $R_0$ and do not diverge 
as $\lambda_\text{th}$.
We obtain excellent agreement of the probability
distribution of the hyperradius $R$ with its analytically known distribution
$\phyp (R)$ (see \fig{snapshot}{\bf d}). This effectively validates our algorithm.
Furthermore, the observed quadratic divergence of the pair distance distribution
 $\ppair(r)$, leading to an
asymptotically constant $r^2 \ppair(r) $ for $r \to 0$, checks with the
Bethe--Peierls condition for the zero-range unitary potential\cite{Bethe1935}
(see \SI{1}).

In local-density approximation, 
particles experience an effective chemical potential $\mu(r) =
\mu_0 - m \omega^2 r^2/2$ that depends on the distance $r$ to the centre of
the trap.  This allows us\cite{Ho2010} to obtain the grand-canonical equation of state 
(pressure $P$ as a function of $\mu$)
from the configurations of a single simulation run at temperature~$T$. 
We find that the equation of state is described very accurately
by the virial expansion up to third order in the fugacity $e^{\beta\mu}$ (see
\fig{eq_of_state} and \SI{4})\cite{Castin2013}.
The third-order term is crucial to the description of Efimov physics
as it is the first term at which three-body effects appear\cite{Huang1987}.
It depends explicitly on $T$ and $R_0$.

In the harmonic trap, particles can be in different thermodynamic phases
depending on the distance $r$ from its centre. We monitor the correlation
between the pair distances and the position in the trap, and are able to 
track the creation of a drop of
high-density liquid at $r\sim 0$ (see \fig{pair_correlations}{\bf a-c}). This drop grows
as the temperature decreases. The observed behaviour corresponds to a
first-order normal-gas-to-superfluid-liquid transition. All particles in the
drop are linked through coherent close-by particle switches as in 
\fig{snapshot}{\bf b}, showing
that the drop is superfluid. Deep inside the liquid phase, the Quantum Monte
Carlo simulation drops out of equilibrium on the available simulation times.
Nevertheless, at its onset and throughout the values of $R_0$, 
the peak of the pair correlation
function is located around $10 R_0$, which indicates that the liquid 
phase is of constant density $n_l \propto R_0^{-3}$ (see \fig{pair_correlations}{\bf c}).
At larger values of $R_0$, the density difference between the trap centre and the
outside vanishes continuously, and the phase transition is no longer
seen (see \fig{pair_correlations}{\bf d-f}). Beyond this critical point, 
the peak of the pair correlation also stabilizes around $10R_0$, 
which indicates a cross-over to liquid behaviour (see \fig{pair_correlations}{\bf f}  
and \SI{5} for additional details on the characterization 
of the first-order phase transition).

Our numerical findings suggest a theoretical model for
the competition between the unitary gas in third-order virial
expansion and an incompressible liquid of density $n_l \propto R_0^{-3}$ 
and constant energy per particle $-\epsilon \propto E_t$,
as suggested by ground-state computations for small clusters\cite{Stecher2010}.
For simplicity, we neglect the entropic contributions to
the liquid-state free energy  $ \epsilon\gg TS$, so that 
$F_\text{liquid} \approx -N\epsilon$. 
The phase equilibrium is due to the difference in free energy and in specific
volume at the saturated vapour pressure (see \SI{6}). 
We extend the third-order virial expansion to describe the gaseous phase in 
the region where quantum correlations become important. 
Because the conventional superfluid transition is continuous,
it still conveys qualitative information about the transition into
the superfluid Efimov liquid in that region.
At small $R_0 \to 0$, the coexistence line approaches infinite temperatures, as
the fundamental trimer energy $E_t\propto R_0^{-2}$ diverges. For larger values of 
$R_0$, the density of the liquid decreases and approaches the one of the gas. 
The liquid--gas transition line
ends in a critical point, were both densities coincide. As the liquid is bound
by quantum coherence intrinsic to the Efimov effect, this critical point
must always be inside a superfluid.
The agreement between this approximate theory and numerical calculations 
for the trap centre phase diagram is remarkable (see \fig{phase_diagram}{\bf a}). 
Beyond the critical point, we no
longer observe a steep drop in the density on decreasing the temperature.
We also notice that quantum coherence builds up in the gaseous phase, 
so that only a conventional superfluid transition takes place.
Our numerical results suggests it occurs at a temperature slightly 
lower than that for the ideal Bose
gas\cite{Dalfovo1999}, $T_\text{BEC}^0\approx\hbar\omega(0.94N^{1/3}-0.69)$. 

Our theoretical model also yields a phase diagram 
for a homogeneous system of unitary bosons (see \fig{phase_diagram}{\bf b}),
where, in addition, the conventional superfluid transition is simply
modelled by that of free bosons\cite{Landau1980}.
In absence of a  harmonic trap, only two independent dimensionless 
numbers may be built, $k_BT / \epsilon$, and $P\hbar^3/
\sqrt{m^{3}\epsilon^{5}}$. As a consequence, the phase diagram in these two
dimensionless numbers is independent of the choice of $\epsilon$, that is, of
$R_0$. The scalings of the rescaled pressure with $R_0^5$ and of the
rescaled temperature with $R_0^2$, explain that the triple point appears much
farther from the critical point in the homogeneous phase diagram than 
in the trap.
We expect model-dependent non-universal effects in two regions.
At high temperature $\lambda_\text{th} \ll R_0$, only a classical gas should exist as
the hyperradial 
cutoff is much smaller than the de Broglie thermal wavelength, and therefore
prevents the build-up of quantum coherence. 
At high pressure $P \propto T$, we expect a classical solid phase
driven by entropic effect, as for conventional hard-sphere melting.

In present-day cold atom experiments, unitarity is achieved by tuning the
scattering length to infinity using Feshbach resonances\cite{Chin2010}. In experimental
systems, the observation of thermodynamic states of the unitary Bose
gas is rendered difficult by the same three-body losses although, on the other
hand, these losses serve to characterize Efimov trimers\cite{Gross2009,
Pollack2009, Zaccanti2009}. Notwithstanding these complications, for
long-lived systems where the recombination rate into non-universal deeply bound
states is sufficiently small\cite{Rem2013}, it should be 
possible to measure the virial corrections to the equation of state and to
probe the universal phase diagram.

\section*{Acknowledgements}
We thank S. Balibar, Y. Castin, F. Chevy, B. Rem, C. Salomon, and F. Werner for very
helpful discussions.  W.K.\ acknowledges the hospitality of the Aspen Center
for Physics, which is supported by the National Science Foundation Grant
No.\ PHY-1066293.

\bibliography{efimov_nature.bib}

\begin{thebibliography}{10}
\expandafter\ifx\csname url\endcsname\relax
  \def\url#1{\texttt{#1}}\fi
\expandafter\ifx\csname urlprefix\endcsname\relax\def\urlprefix{URL }\fi
\providecommand{\bibinfo}[2]{#2}
\providecommand{\eprint}[2][]{\url{#2}}

\bibitem{Efimov1970}
\bibinfo{author}{Efimov, V.}
\newblock \bibinfo{title}{Energy levels arising from resonant two-body forces
  in a three-body system}.
\newblock \emph{\bibinfo{journal}{Phys. Lett. B}}
  \textbf{\bibinfo{volume}{33}}, \bibinfo{pages}{563--564}
  (\bibinfo{year}{1970}).

\bibitem{Efimov1971}
\bibinfo{author}{Efimov, V.}
\newblock \bibinfo{title}{Weakly-bound states of three resonantly-interacting
  particles}.
\newblock \emph{\bibinfo{journal}{Sov. J. Nucl. Phys.}}
  \textbf{\bibinfo{volume}{12}}, \bibinfo{pages}{589--595}
  (\bibinfo{year}{1971}).

\bibitem{Efimov1979}
\bibinfo{author}{Efimov, V.}
\newblock \bibinfo{title}{Low-energy property of three resonantly interacting
  particles}.
\newblock \emph{\bibinfo{journal}{Sov. J. of Nucl. Phys.}}
  \textbf{\bibinfo{volume}{29}}, \bibinfo{pages}{1058--1069}
  (\bibinfo{year}{1979}).

\bibitem{Gross2009}
\bibinfo{author}{Gross, N.}, \bibinfo{author}{Shotan, Z.},
  \bibinfo{author}{Kokkelmans, S.} \& \bibinfo{author}{Khaykovich, L.}
\newblock \bibinfo{title}{Observation of universality in ultracold
  $^7\mathrm{Li}$ three-body recombination}.
\newblock \emph{\bibinfo{journal}{Phys. Rev. Lett.}}
  \textbf{\bibinfo{volume}{103}}, \bibinfo{pages}{163202}
  (\bibinfo{year}{2009}).

\bibitem{Pollack2009}
\bibinfo{author}{Pollack, S.~E.}, \bibinfo{author}{Dries, D.} \&
  \bibinfo{author}{Hulet, R.~G.}
\newblock \bibinfo{title}{Universality in three- and four-body bound states of
  ultracold atoms}.
\newblock \emph{\bibinfo{journal}{Science}} \textbf{\bibinfo{volume}{326}},
  \bibinfo{pages}{1683--1685} (\bibinfo{year}{2009}).
\newblock \bibinfo{note}{{PMID:} 19965389}.

\bibitem{Zaccanti2009}
\bibinfo{author}{Zaccanti, M.} \emph{et~al.}
\newblock \bibinfo{title}{Observation of an {Efimov} spectrum in an atomic
  system}.
\newblock \emph{\bibinfo{journal}{Nature Phys.}} \textbf{\bibinfo{volume}{5}},
  \bibinfo{pages}{586--591} (\bibinfo{year}{2009}).

\bibitem{Rem2013}
\bibinfo{author}{Rem, B.~S.} \emph{et~al.}
\newblock \bibinfo{title}{Lifetime of the bose gas with resonant interactions}.
\newblock \emph{\bibinfo{journal}{Phys. Rev. Lett.}}
  \textbf{\bibinfo{volume}{110}}, \bibinfo{pages}{163202}
  (\bibinfo{year}{2013}).

\bibitem{Braaten2006}
\bibinfo{author}{Braaten, E.} \& \bibinfo{author}{Hammer, H.-W.}
\newblock \bibinfo{title}{Universality in few-body systems with large
  scattering length}.
\newblock \emph{\bibinfo{journal}{Phys. Rep.}} \textbf{\bibinfo{volume}{428}},
  \bibinfo{pages}{259--390} (\bibinfo{year}{2006}).

\bibitem{Stecher2010}
\bibinfo{author}{von Stecher, J.}
\newblock \bibinfo{title}{Weakly bound cluster states of {Efimov} character}.
\newblock \emph{\bibinfo{journal}{J. Phys. B}} \textbf{\bibinfo{volume}{43}},
  \bibinfo{pages}{101002} (\bibinfo{year}{2010}).

\bibitem{Pollock1984}
\bibinfo{author}{Pollock, E.~L.} \& \bibinfo{author}{Ceperley, D.~M.}
\newblock \bibinfo{title}{Simulation of quantum many-body systems by
  path-integral methods}.
\newblock \emph{\bibinfo{journal}{Phys. Rev. B}} \textbf{\bibinfo{volume}{30}},
  \bibinfo{pages}{2555--2568} (\bibinfo{year}{1984}).

\bibitem{Krauth1996}
\bibinfo{author}{Krauth, W.}
\newblock \bibinfo{title}{Quantum {Monte Carlo} calculations for a large number
  of bosons in a harmonic trap}.
\newblock \emph{\bibinfo{journal}{Phys. Rev. Lett.}}
  \textbf{\bibinfo{volume}{77}}, \bibinfo{pages}{3695--3699}
  (\bibinfo{year}{1996}).

\bibitem{Navon2011}
\bibinfo{author}{Navon, N.} \emph{et~al.}
\newblock \bibinfo{title}{Dynamics and thermodynamics of the low-temperature
  strongly interacting {Bose} gas}.
\newblock \emph{\bibinfo{journal}{Phys. Rev. Lett.}}
  \textbf{\bibinfo{volume}{107}}, \bibinfo{pages}{135301}
  (\bibinfo{year}{2011}).

\bibitem{Castin2013}
\bibinfo{author}{Castin, Y.} \& \bibinfo{author}{Werner, F.}
\newblock \bibinfo{title}{Le troisième coefficient du viriel du gaz de bose
  unitaire}.
\newblock \emph{\bibinfo{journal}{Canad. J. Phys.}}
  \textbf{\bibinfo{volume}{91}}, \bibinfo{pages}{382--389}
  (\bibinfo{year}{2013}).
\newblock \eprint{arxiv:1212/5512 (English version)}.

\bibitem{Thomas1935}
\bibinfo{author}{Thomas, L.~H.}
\newblock \bibinfo{title}{The interaction between a neutron and a proton and
  the structure of h{\textasciicircum}\{3\}}.
\newblock \emph{\bibinfo{journal}{Phys. Rev.}} \textbf{\bibinfo{volume}{47}},
  \bibinfo{pages}{903--909} (\bibinfo{year}{1935}).

\bibitem{Chin2010}
\bibinfo{author}{Chin, C.}, \bibinfo{author}{Grimm, R.},
  \bibinfo{author}{Julienne, P.} \& \bibinfo{author}{Tiesinga, E.}
\newblock \bibinfo{title}{Feshbach resonances in ultracold gases}.
\newblock \emph{\bibinfo{journal}{Rev. Mod. Phys.}}
  \textbf{\bibinfo{volume}{82}}, \bibinfo{pages}{1225--1286}
  (\bibinfo{year}{2010}).

\bibitem{Ferlaino2011}
\bibinfo{author}{Ferlaino, F.} \emph{et~al.}
\newblock \bibinfo{title}{Efimov resonances in ultracold quantum gases}.
\newblock \emph{\bibinfo{journal}{Few-Body Systems}}
  \textbf{\bibinfo{volume}{51}}, \bibinfo{pages}{113--133}
  (\bibinfo{year}{2011}).

\bibitem{Feynman1982}
\bibinfo{author}{Feynman, R.~P.}
\newblock \emph{\bibinfo{title}{Statistical mechanics: A set of lectures}}
  (\bibinfo{publisher}{Addison-Wesley}, \bibinfo{year}{1982}).

\bibitem{Krauth2006}
\bibinfo{author}{Krauth, W.}
\newblock \emph{\bibinfo{title}{Statistical Mechanics: Algorithms and
  Computations}} (\bibinfo{publisher}{Oxford University Press, {Oxford, Great
  Britain}}, \bibinfo{year}{2006}).

\bibitem{Bethe1935}
\bibinfo{author}{Bethe, H.~A.} \& \bibinfo{author}{Peierls, R.}
\newblock \bibinfo{title}{The scattering of neutrons by protons}.
\newblock \emph{\bibinfo{journal}{Proc. R. Soc. Lond., A}}
  \textbf{\bibinfo{volume}{149}}, \bibinfo{pages}{176--183}
  (\bibinfo{year}{1935}).

\bibitem{Ho2010}
\bibinfo{author}{Ho, T.-L.} \& \bibinfo{author}{Zhou, Q.}
\newblock \bibinfo{title}{Obtaining the phase diagram and thermodynamic
  quantities of bulk systems from the densities of trapped gases}.
\newblock \emph{\bibinfo{journal}{Nature Phys.}} \textbf{\bibinfo{volume}{6}},
  \bibinfo{pages}{131--134} (\bibinfo{year}{2010}).

\bibitem{Huang1987}
\bibinfo{author}{Huang, K.}
\newblock \emph{\bibinfo{title}{Statistical mechanics}}
  (\bibinfo{publisher}{Wiley}, \bibinfo{address}{New York},
  \bibinfo{year}{1987}), \bibinfo{edition}{2nd} edn.

\bibitem{Dalfovo1999}
\bibinfo{author}{Dalfovo, F.}, \bibinfo{author}{Giorgini, S.},
  \bibinfo{author}{Pitaevskii, L.~P.} \& \bibinfo{author}{Stringari, S.}
\newblock \bibinfo{title}{Theory of {Bose-Einstein} condensation in trapped
  gases}.
\newblock \emph{\bibinfo{journal}{Rev. of Mod. Phys.}}
  \textbf{\bibinfo{volume}{71}}, \bibinfo{pages}{463--512}
  (\bibinfo{year}{1999}).

\bibitem{Landau1980}
\bibinfo{author}{Landau, L.~D.} \& \bibinfo{author}{Lifshitz, L.~M.}
\newblock \emph{\bibinfo{title}{Statistical physics}}.
\newblock No.~\bibinfo{number}{5} in \bibinfo{series}{Course of theoretical
  physics} (\bibinfo{publisher}{Butterworth-Heinemann}, \bibinfo{year}{1980}),
  \bibinfo{edition}{3rd} edn.

\end{thebibliography}

\setcounter{figure}{0}

\begin{figure*}
\centering
\includegraphicsn[width=\linewidth]{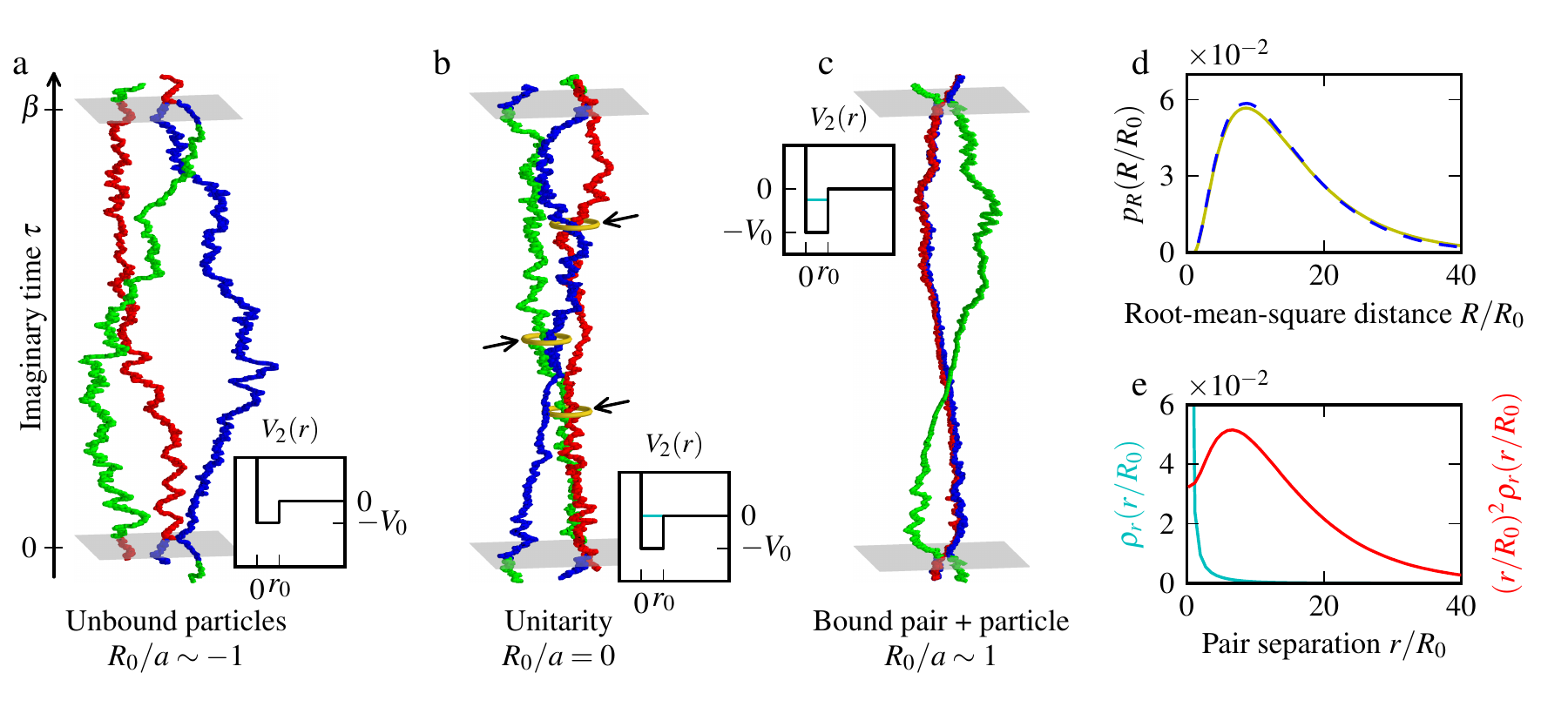}
\caption{\textbf{Path-integral representation of three
  bosons at different scattering lengths} at
  $R_0/\lambda_\text{th}=1.3\times10^{-2}$ (for the graphical projection of $(x, y, z,
  \tau)$ to three dimensions, see \SI{3}). \textbf{a.} At $R_0/a\sim-1$, 
  bosons are unbound and fluctuate on a scale $\lambda_\text{th}$.
  \textbf{b.}  At unitarity ($R_0/a=0$), pairs of bosons form and
  break up throughout the imaginary time (yellow highlights, arrows), forming
  a three-body state bound by pair effects. \textbf{c.} At $R_0/a\sim 1$,
  two bosons bind into a stable dimer \emph{(red, blue)} and one boson
  is unbound \emph{(green)}. Insets in \textbf{a-c} correspond to finite-range
  versions of the zero-range interaction used in each case. Blue levels 
  correspond to the dimer energy. \textbf{d.} Sample-averaged hyperradial
  (root-mean-square pair distance) probability $\phyp(R)$ at constant $\tau$
  \emph{(solid yellow)} compared to its analytic zero-temperature
  value\cite{Braaten2006} \emph{(dashed blue)}. \textbf{e.}~Sample-averaged
  pair distribution $\rho_r(r)$, diverging $\propto 1/r^2$ as $r \to 0$, and
  asymptotically constant $r$-shell probability $r^2 \rho_r(r)$, in agreement 
  with the Bethe--Peierls condition. Data in this figure concern co-cyclic
  particles in a shallow trap of $\omega \sim 0$ (see \SI{3}).  
\flabel{snapshot}} 
\end{figure*}

\begin{figure*}
\centering
\includegraphicsn{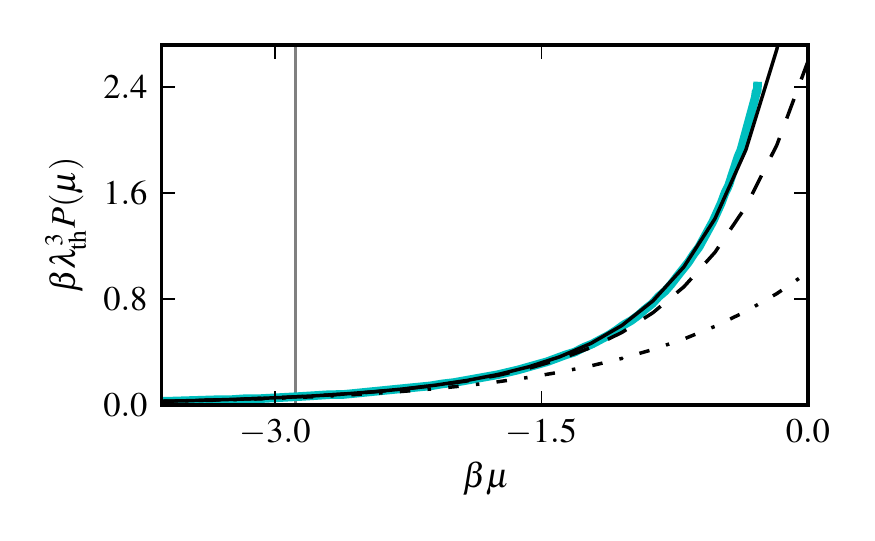}
\caption{\textbf{Equation of state of unitary bosons in a harmonic trap}
  at $R_0/a_\omega=0.18$ and $k_BT/\hbar\omega=38.5$ and $N=100$. The numerical 
  equation of state for $N=100$
  \emph{(solid cyan line)} obtained by ensemble averaging of
  configurations\cite{Ho2010} is compared
  to the theoretical virial expansion up to the 1st \emph{(dash-dotted
  black line)}, 2nd \emph{(dashed black line)}, and 3rd \emph{(solid
  black line)} virial coefficients. The vertical gray line indicates the
  most central region of the trap used to determine $\mu_0$ by comparison
  to an ideal gas.  \flabel{eq_of_state}}
\end{figure*}

\begin{figure*}
\centering
\includegraphicsn{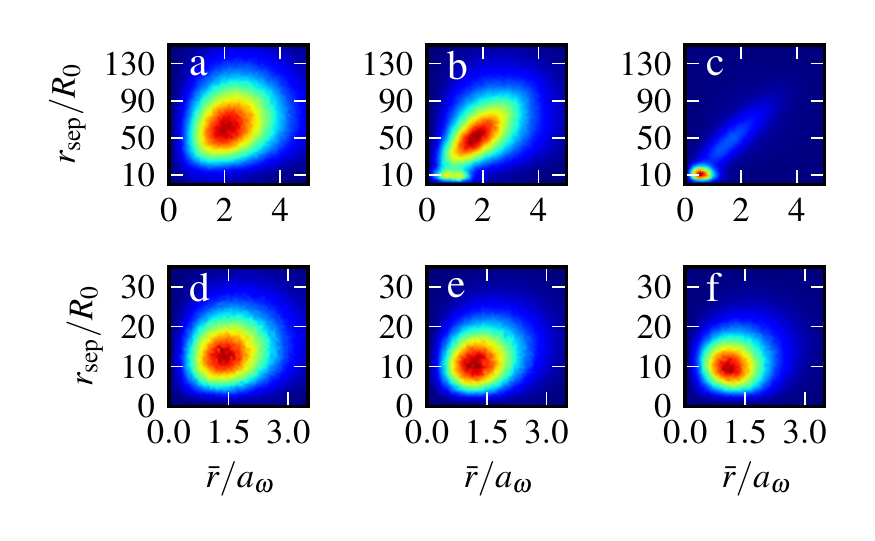}
\caption{\textbf{Two-dimensional histogram of pair distances $r_\text{sep}$ and
  centre-of-mass positions $\bar r$.} \textbf{Upper panels}: 
  \emph{First-order phase transition for $R_0/a_\omega=0.07$}. \textbf{a.}
  At $k_BT/\hbar\omega=6.25$, the distribution is that of the
  normal phase. \textbf{b.} At a slightly lower temperature
  $k_BT/\hbar\omega=5.88$, a second peak with smaller pair distances
  $\sim 10 R_0$ appears in the trap centre. \textbf{c.} At
  $k_BT/\hbar\omega=5.56$,  most particles are in the trap centre, 
  with small pair distances $\sim 10 R_0$. \textbf{Lower panels}: \emph{Smooth
  dependence of pair distances and densities on temperature.} At
  $R_0/a_\omega=0.23$, the pair distances decrease smoothly between
  $k_BT/\hbar\omega=3.33$ \textbf{(d)} and $k_BT/\hbar\omega=2.94$
  \textbf{(e)} and has stabilized around $10R_0$ at $k_BT/\hbar\omega=2.63$
  \textbf{(f)}.  \flabel{pair_correlations}}
\end{figure*}

\begin{figure*}
\centering
\includegraphicsn{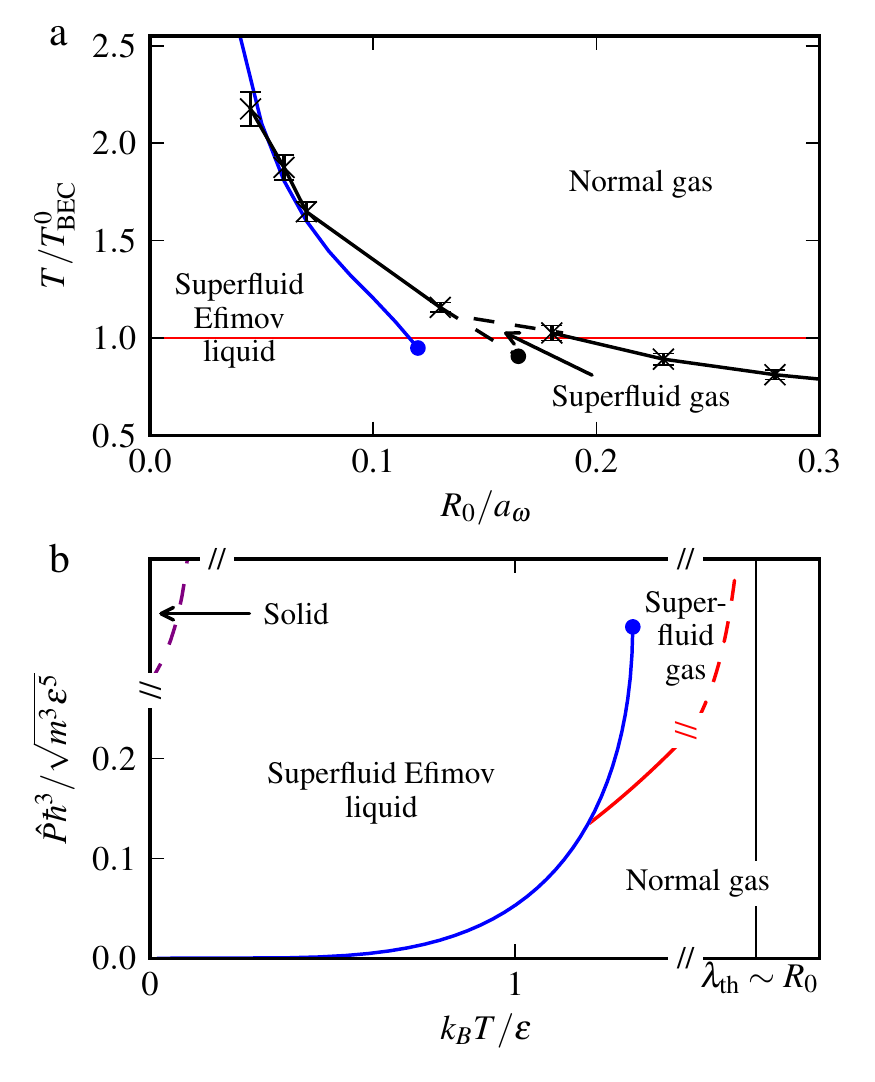}
\caption{\textbf{Phase diagram of the unitary bosons in the trap centre and
in homogeneous space.}
  \textbf{a.} In the harmonic trap, depending on the value of $R_0/a_\omega$, we observe a
  normal-gas-to-superfluid-liquid first order phase transition or
  a conventional second-order superfluid phase transition
  (\emph{solid, black lines}). The normal-gas-to-superfluid-liquid phase
  transition corresponds  well to our theoretical
  model (\emph{solid, blue line}) at high temperatures. Consistently with theoretical
  predictions, the numerical coexistence lines are qualitatively
  continued to a triple point and a critical point \emph{(dashed black
  and dashed green lines)}. 
  \textbf{b.} In a homogeneous system, the normal-gas-to-superfluid-liquid coexistence line \emph{(solid,
  blue)} and the conventional superfluid transition line \emph{(solid, red)} are
  universal. The predicted divergence of the normal-gas-to-superfluid-liquid
  coexistence line \emph{(dashed, red)} and the phase transition to a
  solid phase \emph{(dashed, purple)} are non-universal
  physics specific to our interaction model.  \flabel{phase_diagram}}
\end{figure*}
\end{document}